\documentclass[a4paper,twocolumn,prl]{revtex4}
\usepackage{amsmath}
\usepackage{amsthm}
\usepackage{graphicx}
\usepackage{color}
\usepackage{subfigure}
\usepackage{hyperref}

\newtheorem{theorem}{Theorem}
\newtheorem{corollary}{Corollary}
\newtheorem{definition}{Definition}

\def\a{\alpha}

\def\m{\mu}

\def\n{\nu}
\def\la{\langle}
\def\ra{\rangle}
\def\a{\alpha}

\def\m{\mu}
\def\n{\nu}

\def\h{\hskip 1cm}

\def\lo{\longrightarrow}

\usepackage{color}

\begin{document}

\title{Linking Classical and Quantum Stochastic Processes}

\author {Vahid Karimipour }
\author{Laleh Memarzadeh}
\affiliation{Physics department, Sharif University of Technology, P.O. Box 11155-9161, Tehran, Iran}
\begin{abstract}
We define a map which relates four dimensional classical
stochastic matrices to qubit quantum channels. The map preserves
the spectrum and the composition of processes. To do this we
introduce the concept of Bloch tetrahedron which plays the same
role in the classical context as the Bloch sphere in the quantum
context. A similar map is also induced between dynamical
generators of classical and quantum stochastic processes.
Possibilities for generalization to arbitrary dimensions are also
discussed.
\end{abstract} \maketitle
The state of an open physical system interacting with an
environment on which we have little or no control, is given by a
probability vector $|P\ra$ in the classical domain or a density
matrix $\rho$ in quantum domain. Linear evolution equations
describe the dynamics of such states in successive time steps. In
the classical case, the dynamics is given by
$|P_{n+1}\ra=Q|P_n\ra$ \cite{vankampen} where $n$ denotes the time
step and the stochastic matrix $Q$ collects the jumps
probabilities. In the quantum domain, the dynamics is represented
by a completely positive trace-preserving (CPT) map or quantum
channel $\rho_{n+1}={\cal E}(\rho_n)$ \cite{Kraus}. These
equations provide an abstract but very general framework in which
the dynamics of many different phenomena both in classical and
quantum physics can be analyzed.

In classical domain, this might describe a random walk by a
particle, diffusion and reaction of particles, successive
conformation of a polymer chain, radioactive decay of nuclei, or
population dynamics of any system of interacting entities. Its
application even goes beyond physics to other fields of science,
like biology and finance \cite{Gardiner}.

In quantum domain such an equation may describe an application of
a gate in a quantum processor, a communication channel in time
(quantum storage device) or space (an optical fiber), or any open
systems dynamics where one merely has partial access to the
relevant degrees of freedom \cite{niel}.

How quantum stochastic processes are related to classical ones?
Is it possible to construct a quantum channel for every
stochastic matrix? How about the converse? If yes, how the
properties of the classical stochastic process, are reflected in
the quantum channel? What does it imply about the true quantum
nature of a quantum channel? What types of quantum stochastic
processes can be simulated by classical computers? What are the
precise obstacles for establishing a one-to-one correspondence
between these two domains, at least in those cases where we have
a complete characterization of quantum channels, like the qubit
channels? In this letter we provide results which shed light on
these questions. The point of view adapted in this letter,
complements many other attempts \cite{manyreferences}
for understanding various aspects of
open quantum system dynamics. Specifically we show that\\

a) Four dimensional probability vectors can be characterized in
complete analogy with qubit density matrices, where here Bloch
tetrahedron replaces Bloch sphere,\\

b) four dimensional stochastic matrices can be characterized in
almost complete analogy with qubit channels,\\

c) from these two results we make a map between four dimensional
stochastic matrices and qubit channels, a map which relates many
of the properties, i.e. spectrum,
divisibility etc. of one domain to the other.\\

d) We also show that the map can be extended to relate $d^2$
dimensional stochastic matrices to $d-$ dimensional quantum
channels. This map can play a useful
role in characterization of quantum channels in arbitrary dimension, where a complete characterization is missing.\\

It is worth noting that we go beyond the existing relation
\cite{wolfReferee} between $d$ dimensional stochastic matrices and
the restriction of quantum channels on $d-$ dimensional diagonal
density matrices (In these cases, the image of a stochastic
matrix is always a zero-determinant channel.) We show that the
natural relation is between $d^2$ probability vectors and
$d$-dimensional density matrices, a relation which is highly
nontrivial and very reach indeed. We start with a detailed
exposition in the $d=2$ case.

\begin{figure}
\centerline{\includegraphics[scale=.24]{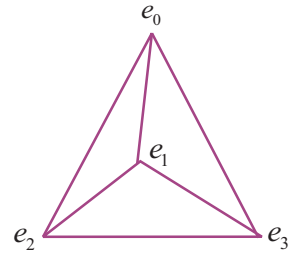}}
\caption{Bloch Tetrahedron: Every vector in this tetrahedron
corresponds to a probability vector according to (\ref{probm}).
The corners correspond to pure states and the center to the
completely mixed state. } \label{fail}
\end{figure}

Consider a system with four configurations. By $P_\m$, $\m=0,\
1,\ 2,\ 3$, we denote the probability of the system to be in each
of these configurations. The state of the system is described by
the vector $|P\ra=\left(\begin{array}{cccc}P_0& P_1 & P_2 & P_3
\end{array}\right)^T$, where $T$ stands for transpose. Obviously
$0\leq P_\m\leq 1$ and $\sum_\m P_\m=1$. So there are three
independent parameters in $|P\ra$. The constraint $\sum_{\m=0}^3
p_\m = 1$ is satisfied if we re-parameterize $|P\ra$ in the
following form
\begin{equation}\label{rep}
  |P\ra=\frac{1}{4}\left(\begin{array}{c} 1+r_1+r_2+r_3\\ 1+r_1-r_2-r_3\\ 1-r_1+r_2-r_3\\1-r_1-r_2+r_3
  \end{array}\right),
\end{equation}
where $r_i$ are three real numbers, the domain of which will be
determined later. The components of $|P\ra$ can be written as
\begin{equation}\label{probm}
  P_\m = \frac{1}{4}(1+e_\m\cdot r),
\end{equation}
where the four 3-dimensional vectors $e_\m$ ($\m=0,1,2,3$)
\begin{eqnarray}\label{em}
&&e_0:=(1,\ 1,\ 1)\ \ \ e_1:=(1,-1,-1)\cr &&e_2:=(-1,1,-1)\ \ \
e_3:=(-1,-1,1),
\end{eqnarray}
point to the four corners of a tetrahedron which we call the
Bloch Tetrahedron and denote it by $\Delta$, figure (\ref{fail}).
These vectors have the following inner products,
\begin{equation}\label{inner}
  e_\m\cdot e_\n = 4\delta_{\m,\n}-1.
\end{equation}
Moreover they satisfy
\begin{equation}\label{outer}
\sum_\m e_\m=0,\h \sum_{\m}(e_\m)_i(e_\m)_j=4\delta_{ij}.
\end{equation}
From these 3-dimensional vectors we can construct an orthornormal
basis for the four dimensional space
\begin{equation}\label{fourvect}
|e_\m\ra=\frac{1}{2}(1,e_\m)^T,
\end{equation} and write $|P\ra$ as
\begin{equation}\label{Pines}
  |P\ra=\frac{1}{2}(|e_0\ra+r_1|e_1\ra+r_2|e_2\ra+r_3|e_3\ra).
\end{equation}

In order that $|P\ra$ be a valid probability vector, all its
entries should be non-negative and less than unity. This yields
the following conditions for the vector $r$,
\begin{equation}\label{condition-r}
  -1\leq e_\m\cdot r\leq 3,
\end{equation}
which means that the vector $r$, should lie in the Bloch
tetrahedron $\Delta$.  In fact $\Delta$ plays the same role for
classical probability vectors as Bloch sphere plays for density
matrices. There is a one-to-one correspondence between classical
probability vectors and points in the Bloch tetrahedron. The
point $r=0$ corresponds to the completely mixed state, where all
configurations are equally probable. The corners of the
tetrahedron correspond to pure states, where only one
configuration has non-zero probability.

The basis $\{|e_\m\ra\}$ allows us to characterize a four
dimensional stochastic matrix in an elegant way, which is
structurally very similar to the characterization of qubit
channels, albeit with important differences which will be
discussed later on. A stochastic matrix $Q$ has the property that
the entries in each column sum up to $1$, that is $\la e_0|Q=\la
e_0|$, hence the expansion
\begin{equation}\label{Qmn}
  Q=|e_0\ra\la e_0| + t_i|e_i\ra\la e_0| + \Lambda_{ij}|e_i\ra\la
  e_j|,
\end{equation}
where a summation over repeated indices is understood. Note that
Greek indices run from $0$ to $3$ and the Latin indices from $1$
to $3$. We denote it by $Q_{t,\Lambda}$, where $t$ is a three
dimensional vector and $\Lambda$ is a real three-dimensional
square matrix. When acting on a probability vector, this matrix
induced an affine transformation on the vector $r$, that is,
\begin{equation}\label{}
  Q_{t,\Lambda}: |P(r)\ra\lo |P(\Lambda r+t)\ra.
\end{equation}
 Therefore $Q_{t,\Lambda}$ will be a
stochastic matrix provided that the affine transformation $
  r\lo r':=\Lambda r + t $
maps Bloch tetrahedron to Bloch tetrahedron (We remind the reader
of a similar requirement of positive maps for which the affine
transformation should map Bloch sphere to Bloch sphere). This puts
severe constraint on the matrix $\Lambda$ and the translation
$t$. For simplicity, hereafter we restrict ourselves to doubly
stochastic marices, for which $t=0$. In order to find the
conditions on the matrix $\Lambda$, we first try to put
$Q_{0,\Lambda}$ into a normal form. \\

\textbf{Normal form of Q:} Consider a doubly stochastic matrix
$Q_{0,\Lambda}$. We can make a singular value decomposition of
$\Lambda$ as
\begin{equation}\label{defnormal}
\Lambda =\hat{S}\Lambda_D \hat{T},
\end{equation}
where the diagonal matrix $\Lambda_D={\rm diagonal}(\lambda_1,
\lambda_2,\lambda_3)$ contains singular values of $\Lambda$.
$\hat{S}$ and $\hat{T}$ are three dimensional orthogonal matrices
and their embedding in four dimensions, in the basis
$\{|e_\m\ra\}$ is of the form
\begin{equation}\label{ST}
   S=\left(\begin{array}{cc} 1 & 0 \\ 0 & \hat{S}
  \end{array}\right)\h T=\left(\begin{array}{cc} 1 & 0 \\ 0 & \hat{T}
  \end{array}\right).
\end{equation}
We will then have
\begin{equation}\label{}
   Q_{0,\Lambda}= S Q_{0,\Lambda_D}T,
\end{equation}
where $Q_n:=Q_{0,\Lambda_D}$ in the right hand side is called the
normal form of the stochastic matrix $Q$ and has the following
expression
\begin{equation}\label{Qmn}
  Q_n:=|e_0\ra\la e_0| + \lambda_{i}|e_i\ra\la
  e_i|.
\end{equation}

Let us now ask under what conditions, the normal form $Q_n$ is a
stochastic matrix. This is
investigated in the next theorem.\\
\begin{theorem}\label{T1}
A matrix $Q_n$ is a doubly stochastic matrix
if and only if the vector $
\lambda:=(\lambda_1,\lambda_2,\lambda_3)$ belongs to $\Delta$. \\
\end{theorem}
\begin{proof} Using (\ref{fourvect}), we find that the explicit form of the
matrix elements of $Q_n$. They are all of the form
$\frac{1}{4}(1+e_\m\cdot \lambda)$ for some $e_\m$. Therefore the
elements of $Q_n$ are positive if and only if $-1\leq\lambda\cdot
e_\m\leq 3,\ \forall \ \m$, which
means that $\lambda\in \Delta$.
\end{proof}
The above representation, as we will see later on, is very apt
for constructing a map between classical four dimensional
stochastic matrices and qubit channels. To this end consider the
following two-dimensional matrices
\begin{equation}\label{A}
A_{\mu}:=\frac{1}{2}(I+e_\m\cdot \sigma),\ \ \ \m=0,1,2,3.
\end{equation}
These matrices are Hermitian, have unit trace and form a basis for
the space of $2\times 2$ matrices, but are not positive. Moreover
they satisfy
\begin{equation}\label{}
  tr(A_\m A_\n)=2\delta_{\m,\n},\h \sum_\m A_\m = 2I.
\end{equation}

\begin{definition}Consider now the following map, $Q\lo {\cal
E}_Q$, which takes a matrix $Q$ to a qubit map
\begin{equation}\label{EqDef}
  {\cal E}_Q(\rho):=\frac{1}{2} \sum_{\m,\n} Q_{\m,\n}tr(\rho
  A_\n)A_\m.
\end{equation}
\end{definition}
This map has the following properties:\\

i) It respects composition, that is
\begin{equation}\label{Homomorphism}
{\cal E}_{Q_1Q_2}={\cal{E}}_{Q_1}{\cal E}_{Q_2},
\end{equation}

ii) It maps an orthogonal matrix like $S$ to a unitary channel,
i.e.
\begin{equation}\label{eses}
  {\cal E}_S(\rho)=U^{-1}\rho U,
\end{equation}
where $U$ is the representation of the orthogonal matrix $S$ on
the basis $A_\m$, i.e. $UA_\m U^{-1}=S_{\m,\n}A_\n$, \\

iii) If $\sum_{\mu=0}Q_{\m,\n}=1$ ($\sum_{\nu=0}Q_{\m,\n}=1$),
then ${\cal E}_Q$ is trace-preserving (unital), finally\\

iv) It preserves the spectrum, that is $Q|v\ra=v|v\ra$, if and
only if ${\cal E}_Q(X_v)=v X_v$, where $X_v:=\sum_{\a}v_\a
A_\a$.\\

From (i), (ii) and the decomposition $Q=SQ_nT$, we find that
\begin{equation}\label{Eq}
  {\cal E}_Q(\rho)= U^{-1} {\cal E}_{Q_n}(V^{-1}\rho V)U.
\end{equation}
Note that $V$ is defined similarly to $U$, i.e. $VA_\m
V^{-1}=T_{\m,\n}A_\n$. Equation (\ref{Eq}) shows the standard
decomposition of any CPT map for qubits \cite{ruskai}. Using
(\ref{A}) and (\ref{EqDef}), the explicit form of ${\cal E}_{Q_n}$
\cite{tozih} turns out to be
\begin{equation}\label{EQn}
  {\cal E}_{Q_n}(\rho)=\frac{1}{2}(tr(\rho)I + \lambda_i tr(\rho \sigma_i) \sigma_i)
\end{equation}

Despite the non-positivity of the matrices $\{A_\m\}$ we have the
following\\
\begin{theorem}\label{T2}
The map ${\cal E}_{Q_n}$ is a unital quantum
channel, if only if $Q_n$ is a doubly stochastic matrix.\\
\end{theorem}

\begin{proof}
The Choi-Jamiolkowski \cite{choi} matrix of ${\cal
E}_{Q_n}$ is found to be
\begin{equation}\label{choi_EQn}
  \tau({\cal E}_{Q_n})=\frac{1}{4}(I\otimes I +  \lambda_i \sigma_i\otimes
  \sigma_i^T).
\end{equation}
The eigenvalues of this matrix can easily be calculated and can
be expressed as
\begin{equation}\label{}
  \frac{1}{4}(1+\lambda\cdot e_\m),\ \ \ \m=0,1,2,3.
\end{equation}
Therefore we have arrived at the nice result that the eigenvalues
of the Choi-Jamiolkowski matrix of ${\cal E}_{Q_n}$ are exactly
the matrix elements of the doubly stochastic matrix $Q_n$. Hence,
these eigenvalues are non-negative if and only if the entries of
$Q_n$ are non-negative. Therefore $Q_n$ is a doubly stochastic
matrix, if and only if ${\cal E}_{Q_n}$ is a completely positive
unital map.\end{proof} Moreover in view of (\ref{Eq}) we find
that the map ${\cal E}_{Q}$ corresponding to $Q\equiv SQ_nT$, is
also CPT and unital, even though for some parameters, the matrix
$SQ_nT$ may no longer be stochastic, since $S$ and $T$ are not
necessarily stochastic.

As a very simple example, the classical depolarizing matrix $
  Q=p\ I + (1-p) |e_0\ra\la e_0|$
is already in normal form and is mapped to the quantum depolarizing channel ${\cal E}_Q(\rho)=p\rho+(1-p)\frac{I}{2}$.\\

The above results suggest that we may also be able to map the
generators of classical stochastic matrices in four dimensions to
Lindblad generators \cite{Lind, breuerbook} of Markovian quantum
evolution for qubits. This is indeed the case, provided that the
matrix generates symmetric stochastic processes, i.e.
$Q_{\m,\n}=Q_{\n,\m}$.

Let $H$ be such a symmetric generator which, due to the condition
$\la e_0|H=0$, will have the following expression in the basis
(\ref{fourvect})
\begin{equation}\label{}
H = \sum_{i,j}H_{ij}|e_i\ra\la e_j|,
\end{equation}
where $H_{ij}=H_{ji}$. In the computational or more aptly the
configurational basis, the off-diagonal elements of such a
matrix, representing the rates of jumps between different
configurations are non-negative and consequently the diagonal
entries are negative. It also has a normal form as
$H=S'H_{n}S'^T$, where
\begin{equation}\label{}
H_n = \sum_ih_i|e_i\ra\la e_i|,
\end{equation}
and $h_i$ are the singular values of $H$. $S'$ is of the same
form as in (\ref{ST}). We now have

\begin{theorem}\label{T3}
The matrix $H_n$ is a generator of a
classical stochastic process, if and only ${\cal E}_{H_n}$ is the
Lindblad generator.
\end{theorem}
\begin{proof}The off-diagonal entries of $H_n$ are all of the
form $h\cdot e_i$ (for $i=1,2,3$), where $h:=(h_1,h_2,h_3)$ and
the diagonal elements are all equal to $h\cdot e_0$. A similar
calculation which led to (\ref{Eq}), now leads to
\begin{equation}\label{ehn}
  {\cal E}_{H_n}(\rho)=\frac{1}{2}\sum_{i}h_i\ tr(\rho \sigma_i)\sigma_i.
\end{equation}
We now resort to a theorem in \cite{assess} according to which,
for a map $\rho\lo L(\rho)$, $L$ is a Lindblad generator if and
only if it is Hermitian, ${L}^*(I)=0$, where $L^*$ denotes the
dual map, and $\omega^{\perp}\hat{L}^{\Gamma}\omega^{\perp} \geq
0$. In this last expression, $\omega^{\perp}=I-|\omega\ra\la
\omega|$ where $|\omega\ra=\frac{1}{\sqrt{2}}(|00\ra+|11\ra)$ is a
maximally entangled state, and $\hat{L}$ is the linearized form
of the map, when it acts on the vectorized form of $\rho$,
$\rho_{i,j}\lo \hat{L}_{ij,kl}\rho_{k,l}$ and finally
${(.)}^\Gamma$ represents the involution
$(\hat{L}^{\Gamma})_{ij,kl}=\hat{L}_{ik,jl}$. The first two
conditions are obviously satisfied. For the third we note from
(\ref{ehn}) that ${\hat{\cal E}_{H_n}}^{\Gamma}=\frac{1}{2}\sum_i
h_i (\sigma_i\otimes \sigma_i^t)$ that the eigenvalues of
$\omega^{\perp}{\hat{\cal E}_{H_n}}^{\Gamma}\omega^{\perp}$ are
exactly equal to ${{\frac{1}{2}}} h\cdot e_i$, which are nothing
but the off diagonal elements of $H$. So ${\cal E}_{H_n}$ is a
quantum Lindblad generator if and only if $H_n$ is a classical
generator.
\end{proof}
\begin{corollary}\label{coro1}
For any matrix of the form $H=S'H_nS'^T$,
where $S'$ is as in (\ref{ST}), the corresponding ${\cal E}_H$
will also be a Lindblad generator.
\end{corollary}
\begin{proof}Let $U$ be as before, (i.e. after equation
(\ref{eses})). Then it is clear that
\begin{equation}\label{}
  {\cal E}_H (\rho)=U^{-1}  {\cal E}_{H_n} (U\rho U^{-1})  U,
\end{equation}
Using the relation $\hat{\cal E}^\Gamma=2\tau({\cal E})$ we find
\begin{equation}\label{replacement}
{{\hat{\cal E}^\Gamma_H =(U\otimes U^*)^{-1}  \hat{\cal
E}^\Gamma_{H_n} (U\otimes U^*).}}
\end{equation}
Now since $\omega^\perp$ commutes with $U\otimes U^*$, we find
$\omega^\perp{\hat{{\cal E}_{H_n}}}^\Gamma\omega^\perp\geq 0$ if
and only if $\omega^\perp{\hat{{\cal
E}_{H}}}^\Gamma\omega^\perp\geq 0$. Therefore the six-parameter
family of the matrices $H$ also lead to quantum Lindblad
generators. In such a case, if $Q=e^H$, then we find that ${\cal
E}_Q=e^{{\cal E}_H}$, that is $H$ and ${\cal E}_H$ generate
respectively
the classical stochastic process and its quantum image. \end{proof}
Returning to our map $Q\lo {\cal E}_Q$, we would like to
elaborate on the decomposition of stochastic matrices and the
effect it has on their quantum counterpart. Consider the normal
form $Q_n$. For $\lambda$ to lie inside the tetrahedron, the
magnitude of components $\lambda_i$ should be less than one. The
matrix $Q_n$ shrinks the Bloch Tetrahedron in the principal
directions by a factor $\lambda_i$. The transformed tetrahedron
still lies inside $\Delta$. However rotations and reflections
induced by $S$ and $T$ will bring small parts of the new
transformed tetrahedron to outside $\Delta$. The reason is that
the matrices $S$ and $T$ are only orthogonal and not stochastic
\cite{estesna}. Therefore while $SQ_nT$ may be non-stochastic, its quantum counterpart ${\cal U}\circ {\cal E}_{Q_n}\circ {\cal V}$ is certainly a qubit channel.  \\
Conversely there may be stochastic matrices for which $\lambda$
does not belong to  $\Delta$, whose quantum image is not a qubit
channel. As expected, there shouldn't be a one-to-one
correspondence between classical and quantum processes,
nevertheless we have found a map from stochastic matrices, whose
image has a very large overlap with qubit channels and have also
identified why such an overlap is not complete. It seems that two
types of mapping are possible, one can either use orthnormal and
hence non-positive operators, (like
$A_\m:=\frac{1}{2}(I+e_\m\cdot \sigma)$ in $d=2$) or else choose
positive and non-orthonormal operators, (like
$A_\m:=\frac{1}{2}(I+\frac{1}{\sqrt{3}}e_\m\cdot \sigma)$). We
think that these two kinds of maps complement each other, and
they are induced from two types of mapping the Bloch tetrahedron
to the Bloch sphere, a subject which we will explore in more
detail elsewhere \cite{kmm}.

To what extent these considerations can be generalized? Let
$\{A_\m\}$ and $\{B_\m\}$ be two basis sets of Hermitian
operators in $M_{d^2}$, (the space of $d^2\times d^2$ matrices)
with the following properties:
\begin{equation}\label{AB}
  \sum_\m {A_\m}=I, \ \ \
  tr(B_\n)=1.
\end{equation}
We can then define a map $\Phi$, which takes any $Q\in {M}_{d^2}$
to an ${\cal E}_Q$ defined as
\begin{equation}\label{map}
  {\cal E}_Q(\rho):=\sum_{\m,\n} Q_{\m,\n} tr(\rho A_\n) B_\m.
\end{equation}
If the operators $\{A_\m\}$ and $\{B_\n\}$ are positive, (which
will no longer be orthonormal), then $\Phi$ maps any stochastic
(and doubly stochastic) matrix to a CPT (and unital) map, readily
verified by forming the Choi-Jamiolkowski matrix of ${\cal E}_Q$.
It also follows the composition rule
\begin{equation}\label{}
  {\cal E}_{Q_1}{\cal E}_{Q_2}={\cal E}_{Q_1GQ_2},
\end{equation}
where $G$ is the stochastic matrix given by $
  G_{\m,\n}=tr(A_\m B_\n)$.

In summary we have suggested a way for mapping classical $d^2$
stochastic matrices to quantum channels in $d$ dimensions. In
particular, in $d=2$, we have shown that four dimensional
probability vectors and stochastic matrices, can be described in
almost complete analogy with qubit density matrices and qubit
quantum channels. This similar structure allows us to make a
one-to-one correspondence between the normal form of doubly
stochastic matrices in four dimensions and the normal form of
qubit quantum channels. These findings may be expanded in a few
directions. Besides considering the case of non-doubly stochastic
matrices, it obviously begs for a generalization to arbitrary
dimensions, a task which is highly non-trivial. The immediate
task is to generalize it to the case of multi-qubits, where a
uniformly spaced basis like (\ref{fourvect}) exists due to the
existence of Hadamard matrices in $2^k$ dimensions. Once this is
done, one can also see how classical correlations lead to
correlated quantum channels.  Moreover these findings and the
homomorphic property (\ref{Homomorphism}), may shed light on the
problem of divisibility and Markovianity
\cite{manyreferences} of
classical and quantum maps and relate this problem in these two domains.\\

\textbf{Acknowledgements:} V. K. would like to thank the Abdus
Salam ICTP, where part of this work was done. We would like to
thank A.
T. Rezakhani, M. R. Koochakie, M. H. Zare and A. Mani for valuable comments.\\


\begin{thebibliography}{}
\bibitem{tozih} For this one should also use the form
$(Q_n)_{\m,\n}=\frac{1}{4}(1+\lambda_i(e_\m)_i(e_\n)_i)$ and the
properties (\ref{outer}).
\bibitem{estesna} When $\lambda\in S(\frac{1}{\sqrt{3}})$, the latter being the
sphere inscribed in the Bloch tetrahedron, then
$SQ_{0,\lambda}T$  is a doubly stochastic matrix for all
orthogonal matrices $S$ and $T$ of the form (\ref{ST}).
\end{thebibliography}
\end{document}